\documentclass{PoS}

\title{Heavy Flavours in DIS and Hadron Colliders:\\ Working Group Summary}

\ShortTitle{Heavy Flavours in DIS and Hadron Colliders}

\author{Katerina Lipka\\
        DESY\\
        Notkestrasse 85, D-22607, Hamburg, Germany\\  
        E-mail: \email{katerina.lipka@desy.de}}

\author{Gennaro Corcella\vspace{0.1cm}
\\
 Museo Storico della Fisica, Centro Studi e Ricerche E.~Fermi,\\
             Piazza del Viminale 1, I-00184 Roma, Italy,\\   
Scuola Normale Superiore, Piazza dei Cavalieri 7, I-56123 Pisa, Italy,\\
INFN, Sezione di Pisa,  Largo Fibonacci 3,
I-56127, Pisa, Italy\\
        E-mail: \email{gennaro.corcella@sns.it}}

\abstract{
The recent theory developments and latest experimental results on heavy-flavour production in Deep 
Inelastic Scattering and at hadron colliders are summarized. Models of heavy quarkonia production, 
non-perturbative corrections to fragmentation, theory of heavy-hadron production in heavy-ion collisions, 
and interpretation of new exotic hadrons are discussed. Progress in event generators development is 
reported. Most recent experimental results from HERA and $e^+e^-$ colliders as well as from proton-(anti)proton 
and heavy ion experiments are presented and the role of charm and beauty quarks in
the analyses of the proton structure is stressed.}

\FullConference{XVIII International Workshop on Deep-Inelastic Scattering and Related Subjects, DIS 2010\\
		April 19-23, 2010\\
		Firenze, Italy}

\begin{document}

\section{Introduction}
Heavy flavours (charm, bottom and top quarks) play a particularly important role in particle physics and
their production in different environments is interesting from both theoretical and experimental points of view. 
The large masses of charm, beauty and top quarks with respect to the hadronisation scale, 
say $\Lambda_{\mathrm{QCD}}^{\overline{\mathrm{MS}}}\approx$~200-300~MeV, makes perturbative QCD 
(pQCD) applicable.
 
Fixed-order QCD calculations are reliable to calculate total heavy-quark production cross sections. However, 
resummations are necessary in order to improve differential distributions, since those typically exhibit 
contributions which are enhanced for soft- or collinear-parton radiation and need to be summed to all orders. 
Also, perturbative QCD is unable to describe the transition of a heavy quark into a heavy hadron,
but a reliable hadronisation model is needed.

Studying heavy-flavour production and decay provides precision tests of QCD and of the Standard Model of
the electroweak interactions. Such, in lepton-nucleon scattering, the charm- and beauty-production 
mechanism is directly sensitive to parton distributions in the nucleon. Production of heavy quarks is a 
tool to test the Standard Model in electron-positron collisions and a unique probe of the properties 
of the
hot and dense medium created in heavy-ion experiments. Being the heaviest fundamental particle, 
the top quark 
continues to be an exciting topic for both experiment and phenomenology. 
The determination of the top quark  
properties provides a perfect testing ground of the Standard Model: in fact, 
the top quark plays a crucial role in 
the electroweak precision tests and its mass constrains the Higgs mass.

Modern heavy-flavour physics represents an enterprise of precise experimental measurements 
and theory developments connected in a common effort. The most recent results achieved by experiments 
and theory groups since last DIS workshop in 2009 were lively discussed in the working group. In the following,
the main issues are summarised. Special attention is given to the treatment of heavy quarks in the QCD 
analyses of the proton structure functions, discussed in a joined session 
between heavy-flavour and PDF working groups.

\section{Theory Issues}

The presentations on the theory and phenomenology of heavy flavours dealt with heavy-quark 
and heavy-flavoured hadron production in DIS, $pp$, $p\bar p$, $e^+e^-$ and heavy-ion collisions, 
top quark phenomenology and the observation of exotic hadrons at the Tevatron.

\subsection{$J/\psi$ photo- and electroproduction}

New results were presented for the photoproduction of $J/\psi$ at
next-to-leading order (NLO) in the framework of Non Relativistic
Quantum Chromodynamics (NRQCD) \cite{kniehl}, which provides a rigorous 
factorisation theorem for heavy-quarkonium production and decay.
The production of $c\bar c$ pairs can be calculated in pQCD, whereas the
transition to the $J/\psi$ involves non-perturbative matrix elements, 
containing parameters which must be fitted to experimental data.
In NRQCD, the amplitudes can be written as a double expansion with
respect to the relative quark-antiquark velocity $v$ and 
the strong coupling constant $\alpha_S$.
The LO term in $v$, i.e. ${\cal O}(v^2)$, corresponds to the 
state $^3S_1^{[1]}$, which is colour-singlet.
At NLO, i.e. ${\cal O}(v^2)$, also the colour-octet states, namely $^3S_1^{[1]}$, $^3S_1^{[8]}$ and
$^3P_{0/1/2}^{[8]}$, should be accounted for, as discussed in \cite{kniehl} for $J/\psi$ photoproduction.
After the inclusion of NLO colour-octet contributions, the calculations describe the HERA measurements of 
the photon-proton invariant mass or $J/\psi$ transverse momentum within the theoretical uncertainties.
However, discrepancies with data are observed in the $J/\psi$ elasticity $z$ distribution,
at small values of $z$.

$J/\psi$ photo- and electroproduction was also discussed in the
framework of a dual model \cite{prokudin}. In fact, the idea of
duality is that, in a scattering process, 
the sum over intermediate resonances in the $s$-channel is
equivalent to Regge exchanges in the $t$-channel. 
The model \cite{prokudin} consists in Pomeron ($P$) exchange and
smooth non-resonant background close to the $J/\psi$ threshold,
i.e. $M_P+M_{J/\psi}$.
It presents two trajectories and a few parameters, namely
8 for photoproduction and 3 more for electroproduction, which must
be tuned to data. 
The predictions of the model on differential and elastic cross sections
at HERA were compared with the data and exhibited an overall good 
agreement for both $J/\psi$ photo- and electroproduction.

\subsection{Heavy quark production and Regge theory}

The study \cite{saleev} discusses 
inclusive hadroproduction of single $b$-jets, $b\bar b$ dijets
and $b\gamma$ dijets at the Tevatron in the Regge limit at leading order.
The calculation is based on an effective field theory, with a non-Abelian
gauge-invariant action, and is characterised by Reggeized quarks and gluons.
Within suitable ranges of possibly ordered rapidities, $b$, $b\bar b$ and $b\gamma$
production can be described
by means of a $t$-channel scattering mediated by
Reggeized quarks and gluons.
In spite of the simplicity of its formulae and assumptions, this model describes 
the Tevatron 
data in the relevant kinematics regimes. In detail, its predictions fare 
rather well with respect to the CDF data on
transverse momentum ($b$ jets), transverse energy, 
invariant mass, pseudorapidity, azimuthal
distributions ($b\bar b$ dijets) and photon transverse momentum ($b\gamma$ dijets).

\subsection{Higher-order QCD calculations for heavy quarks
and extraction of the the gluon polarisation}

Progresses towards the computation of 
the heavy-quark contributions to the structure
function $F_2$, to three-loop accuracy, were reported in \cite{bluemlein}. 
The calculation of the massive Wilson coefficients was carried out
in Mellin moment space for generic $N$ and is based on hypergeometric functions
and advanced summation techniques. The large logarithms
of the ratio of the factorisation scale and heavy-quark mass, i.e.
terms 
$\sim\alpha_S^k\ln^m(\mu_F^2/m^2)$, have already been calculated for $k=3$ and
$m\leq 3$ in $N$-space and should be soon available even in $x$-space.
The status of the calculation of the contribution $\sim T_F^2n_fC_{A,F}$ to the structure
function $F_2(x,Q^2)$ was also presented.

New results on threshold resummation, i.e. large-$x$ contributions, for charm production
in Neutral Current DIS were presented. In particular, as discussed in \cite{lopresti}, the 
coefficient function $C_2$ contains in Mellin space
enhanced logarithms $\alpha_S^n\ln^k N$, with $k\leq 2n$, and logarithms
of the heavy-quark velocity, i.e. $\ln^m\beta$, enhanced for 
$\beta\to 0$.  The charm structure function
$F_2^c (x,Q^2)$ was then computed 
at fixed order, i.e. NLO and NNLO, and including NLL threshold
resummation, employing the ABKM(09) parton distributions. In particular,
Ref.~\cite{lopresti} shows $F_2^c(x,Q^2)$ at fixed order, with the inclusion
terms $\sim\ln\beta$ and $\sim\ln^2\beta$, and accounting for Coulomb corrections
as well. 
Furthermore, the differential distribution $dF_2^c/dp_T^2$,
obtained for fixed $Q^2$ and $x$, exhibits milder dependence on the factorisation
scale once NLL resummation is included.

As far as higher-order calculations are concerned, 
a progress report on the status of the NLO 
corrections to 4 $b$-quark production at the LHC was presented in \cite{greiner}.
In fact, the production of four beauty quarks in $pp$ collisions is particularly
relevant since it is a background
for Higgs decays $H\to b\bar bb\bar b$ in the MSSM.
The calculation of real and virtual diagrams was undertaken by
using the subtraction method for soft and collinear singularities:
the $q\bar q\to 4b$ computation is available, while the full $pp\to 4b$ has not
been completed yet. Preliminary results exhibit a strong
impact of the NLO calculation: the dependence on the renormalisation scale
is much milder and the prediction of the invariant-mass distribution of
a final-state $b$-quark is more stable after the inclusion of the
higher-order terms. It was also pointed out that, although the
uncertainty on the prediction gets much lower, the shapes
of invariant-mass and $b$ transverse-momentum spectra is 
roughly the same, regardless of the NLO contributions.

The extraction of the gluon polarisation from open $D^0$ production at
COMPASS was investigated in~\cite{kurek}.
The asymmetries measured by the COMPASS experiment are 
compared to the predictions of the AROMA event generator,
which includes LO/NLO hard-scattering cross section and parton showers. 
In this way, one can extract the gluon polarisation and distinguish the two cases $\Delta g>0$ 
and $\Delta g<0$.

\subsection{An effective-coupling model for heavy-quark fragmentation}

A novel model to include non-perturbative corrections to heavy-quark production was  
discussed in~\cite{ferrera} 
and applied to $B$ and $D$ production in $e^+e^-$ annihilation at LEP and
$B$-factories. Unlike the conventional
approach, which uses hadronisation
models depending on a few tunable parameters, 
one can construct an effective strong
coupling constant in such a way that a power-suppressed
term subtracts the Landau pole off.
Such an effective coupling is thus
finite at small scales.
Following \cite{ferrera}, heavy-quark production is
described in the framework of perturbative
fragmentation functions, which factorises the heavy-quark spectrum
as a convolution of a massless process-dependent coefficient function
and a process-independent perturbative fragmentation function,
expressing the transition of a light parton into the heavy quark.
The coefficient function is thus calculated at NLO, the perturbative
fragmentation function evolves to NLL accuracy, and 
NNLL threshold resummation is implemented in
both coefficient function and initial condition of the perturbative
fragmentation function.
The effective-coupling model
is able to describe, within the theoretical error, the
$B$- and $D$-hadron energy spectra measured at LEP in $x$-space, whereas
significant discrepancies are present when comparing with
$D$-hadron data from $B$-factories. All considered data,
however, can be reproduced in $N$-space.
The inclusion of further NNLO terms in the coefficient function
and in the Altarelli--Parisi splitting functions should 
decrease the theoretical 
uncertainties and, at the same time, shed light on 
the comparison with $B$-factory data on charmed hadrons, which,
for the time being, cannot be reproduced in $x$-space.

\subsection{Monte Carlo generators for heavy flavour physics}

Progress in the implementation of heavy-quark production and decay
in Monte Carlo event generators was presented.
In POWHEG~\cite{re}, 
NLO calculations are matched with parton cascades without yielding negative weights 
and in a manner which is independent of the particular shower model. 
Thanks to the fact that the first emission is generated
at the largest transverse momentum, this method can be implemented straightforwardly in the
framework of showers ordered in transverse momentum, e.g. the latest PYTHIA version. 
However, 
it would be necessary to include the so-called truncated showers to recover colour coherence, when 
running angular-ordered parton cascades, such as HERWIG.
Top-pair production has been implemented in POWHEG since a while ago
and lately even single-top 
production in both $s$- and $t$-channels has been included. The implementation of the $Wt$ 
channel, i.e. scatterings like $gb\to Wt$, which are negligible at the Tevatron, but
relevant at the LHC, is still in progress. 
Several results were shown in~\cite{re} 
for single-top production at the Tevatron, 
including transverse-momentum, rapidity and angular distributions.
POWHEG predictions are in reasonable agreement with MC@NLO, the other Monte Carlo program 
available to match NLO computations with HERWIG parton showers. 
Moreover, it exhibits remarkable 
improvement with respect to PYTHIA, which includes extra-parton radiation in single-top processes
only in the soft/collinear approximation, as matrix-element corrections are
presently missing. For example, unlike POWHEG, PYTHIA does not generate $B$-hadrons at large
transverse momentum. In addition, POWHEG includes angular correlations between 
the top-decay products 
a posteriori, providing the user with angular distributions, which are not available in PYTHIA.

The CASCADE event generator was also reviewed in~\cite{kramer}: 
initial-state multiple radiation 
is implemented according to the CCFM equation, whereas final-state parton showers satisfy 
the angular-ordering prescription and DGLAP evolution. Also, it is important
to point out that in CASCADE gluons are the only available partons in the proton. 
As for the initial state, 
CASCADE uses 
unintegrated parton distributions, obtained from a fit based on the CCFM equations, using H1 
data for $x<0.005$ and $Q^2>5$~GeV$^2$:
in this way, also non-collinear evolution is included. 
Hadronisation is finally simulated by means of the Lund string model.
Ref~\cite{kramer} presents studies on
the transverse momentum of $D^+$ and $B^+$ hadrons 
at the Tevatron and CASCADE is able to reproduce the CDF data.
Comparisons with data on jet-jet (hadron-hadron) azimuthal separations were also discussed: 
as pointed out in \cite{kramer}, in order to 
reproduce the CDF data, it was necessary to vary the upper limit of the evolution 
variable from $Q^2=m^2$, $m$ being the heavy-quark mass, to $Q^2=m_T^2$, where $m_T$ 
is the transverse mass. 
Good agreement with the measurements 
of $\Delta\Phi_{jj}$ and $\Delta\Phi_{D^0D^+}$ was thus obtained. 
CASCADE was also compared to $D^*$ and jet 
photoproduction at H1 \cite{staykova}, but it turned out to be unable to 
describe the shapes of the double-differential distributions.

\subsection{Open heavy quarks in heavy-ion collisions}

Several issues related to open heavy-flavour production in heavy-ions collisions were 
discussed~\cite{armesto}. 
The FONLL calculation, based on a NLL soft/collinear resummation,
matched to the NLO cross section, works well for $b$ production at the Tevatron,
but lies systematically below the heavy-hadron production measurements of STAR and PHENIX experiments at RHIC.
Nevertheless, the RHIC measurements of
observables relying on semi-leptonic decays of heavy hadrons 
are described well for $p_T>$~2.5 GeV, within the errors.
Progress on the determination of nuclear parton densities was also 
reported: the current nuclear PDF sets are 
available to NLO and are obtained using data from Drell--Yan interactions, $\pi^0$ decays, 
and Neutral Current DIS, whereas the inclusion of neutrino data is under discussion.

The main mechanisms for heavy-quark energy loss in a dense medium
were then reviewed: radiative and collisional energy loss, which do not account for hadronisation, 
and meson dissociation, which is instead 
driven by hadronisation. When comparing with RHIC data, radiative energy loss
tends to overestimate the data, except for charm-quark production. 
Collisional dissociation, including radiative
contributions, is instead able to 
describe the data, as long as one assumes that hadronisation is almost
instantaneous.
Predictions for the LHC were shown~\cite{armesto}: remarkably, 
both collisional and radiative 
scenarios lead to different transverse-momentum spectra for charm and beauty quarks, 
whereas the meson-dissociation model 
predicts similar suppression for $D$ and $B$ hadrons. 
These results imply that at the LHC the separation of
charm and beauty production should be possible.

\subsection{Loosely bound molecules at hadron colliders}

Production of loosely-bound molecules at hadron colliders was addressed
in~\cite{sabelli}, taking particular 
care about the discovery of the $X(3827)$ particle by the CDF 
collaboration at the Tevatron. 
As it is a $J^{PC}=1^{++}$ state, it could not be interpreted as 
standard charmonium, and therefore it was
first described as a $D^0\bar D^{0*}$ molecule with a
radius $r=8$~fm. However, 
using the hadronisation models contained in HERWIG or PYTHIA to 
estimate the transition probability
of $p\bar p$ into a $DD^*$ molecule, one obtains 
a much lower cross section for $X(3827)$ production than the measured one.  
In principle, one may invoke final-state interactions 
and the Watson theorem to reconcile theory and experiment, 
but nonetheless this theorem can be applied only for
two-particle final states, which is not the case in the Tevatron environment.
Also, if the $X(3827)$ were indeed a molecule, CDF should have even
observed a $D_sD^*_s$ molecule, but such an observation was never reported so far.
The conclusion of~\cite{sabelli} is therefore that the hypothesis 
of tetraquarks should possibly be reconsidered. In this case, it will be
particularly interesting searching 
for a $X^{++}$, i.e. a $[cu][\bar d\bar s]$ state, at the LHC in the
next few years.

\section{Experimental Issues}

Recent results on charm and beauty production in lepton-nucleon collisions were 
presented by the H1, ZEUS, NOMAD and COMPASS experiments. 
BaBar and Belle reported on Standard Model tests using heavy quarks in electron-positron 
collisions. STAR and PHENIX experiments discussed
heavy quarks as probes of quark-gluon plasma; CDF and D0 showed 
precision measurements of the top quark properties.
First physics using the LHC data was presented by ALICE, ATLAS, CMS and LHCb 
experiments.  
  
\subsection{QCD tests using heavy-quarkonium production}
 
\begin{figure}[htbp]
\centerline{\resizebox{0.53\textwidth}{!}{\includegraphics{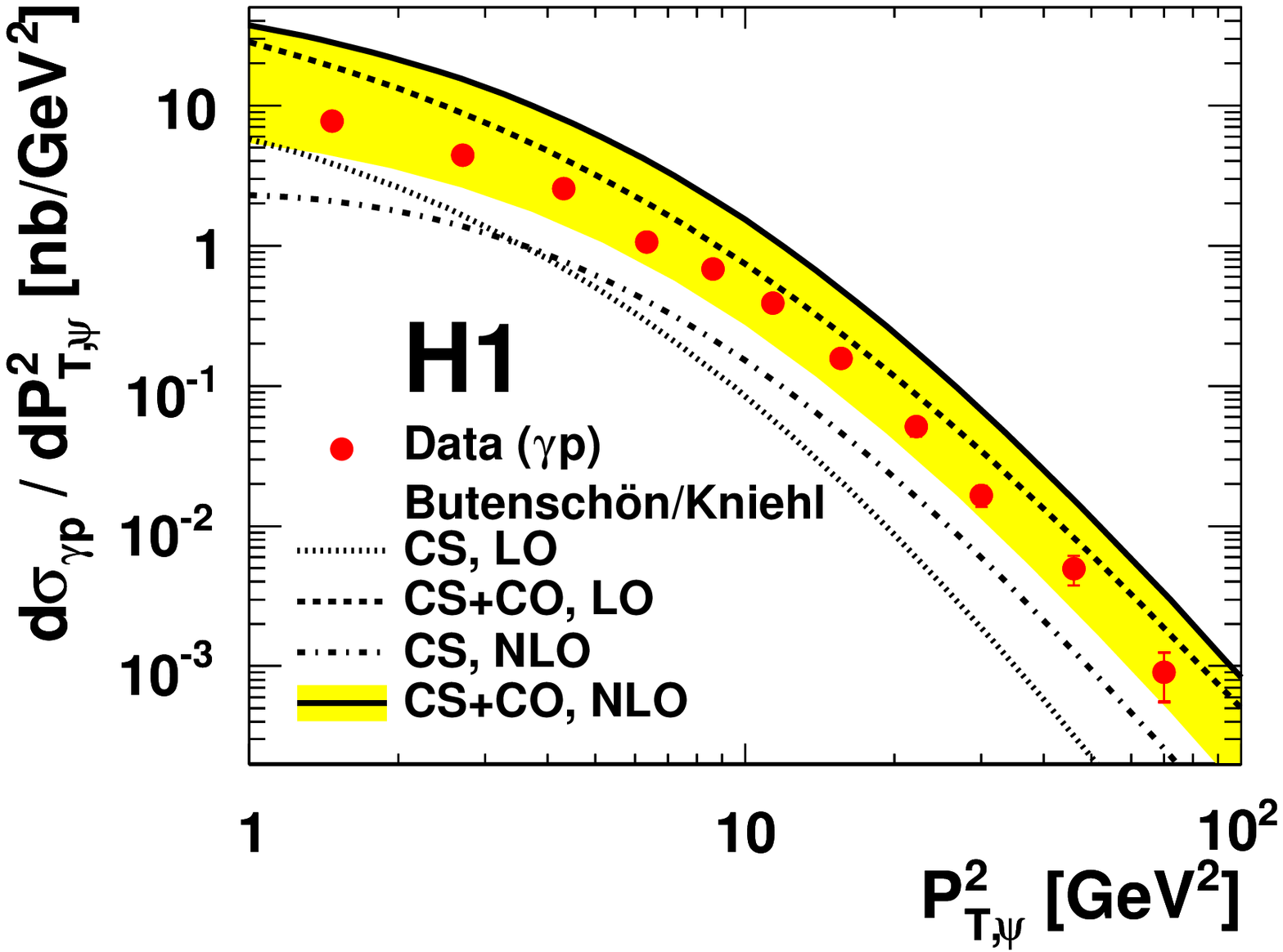}}%
\hfill%
\resizebox{0.46\textwidth}{!}{\includegraphics{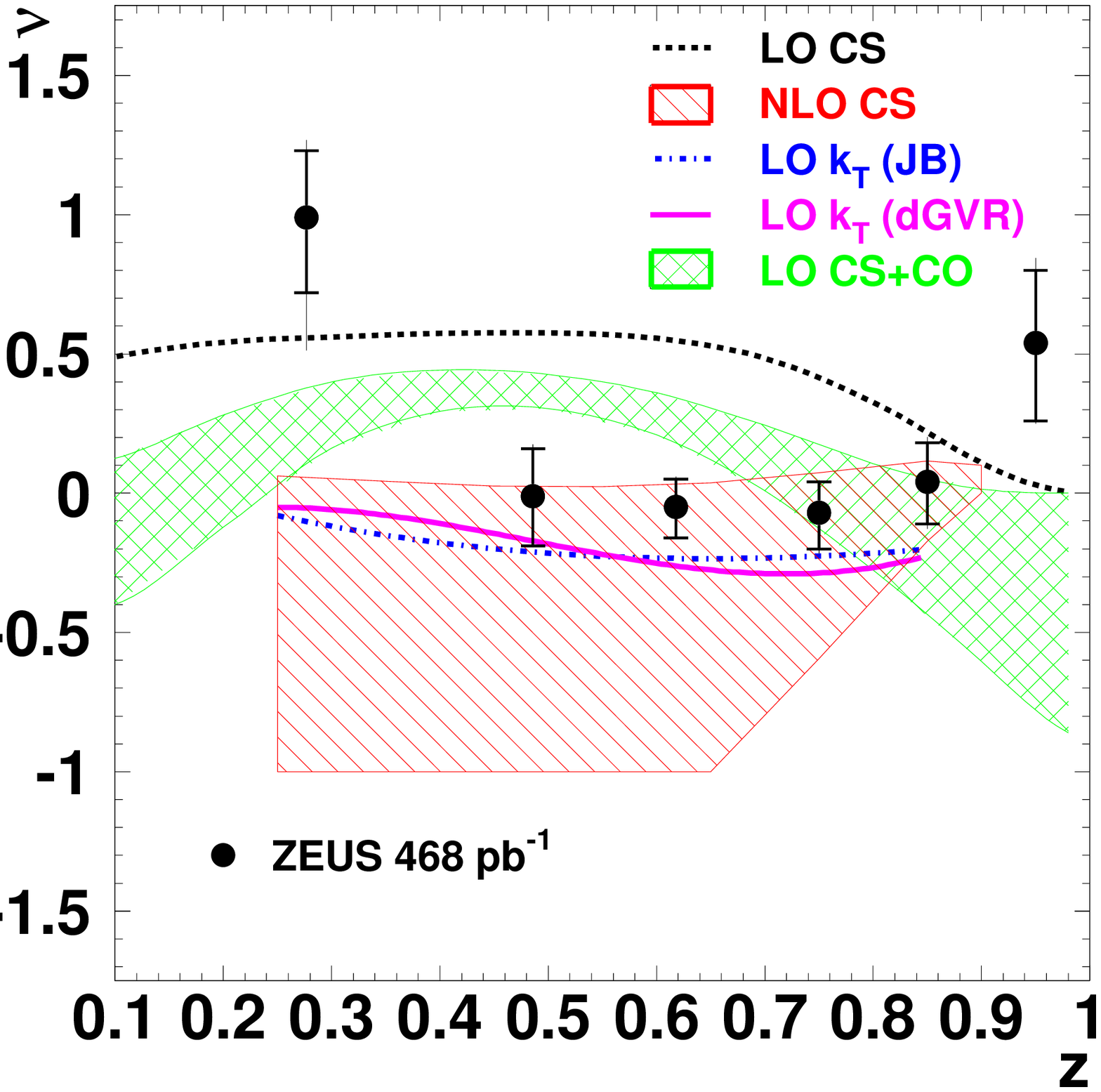}}}
\caption{Left: Cross section of $J/\psi$ photoproduction as a function of 
the meson transverse momentum $p_T$. Right: 
Polarisation parameter $\nu$ as a function of the fractional photon energy $z$ 
carried by the meson.}
\label{fig1}
\end{figure}

Heavy-quarkonium production in $ep$ and $p\bar p$ collisions serves for precision tests of perturbative QCD 
calculations as well as for the modelling of hadronisation corrections. At HERA inelastic $J/\psi$ photoproduction 
cross sections and polarisation are measured \cite{steder,bertolin}  and compared to the calculations of 
Colour-Singlet (CS) and Colour-Octet (CO) models in Non-Relativistic QCD (NRQCD), and in the framework of 
$k_T$-factorisation in Fig.~\ref{fig1}.
The CS model at NLO describes the shape of the $p_T$ distribution quite well, while it is 
unable to predict the overall normalization. After including the CO contribution, the 
NLO NRQCD prediction is able to reproduce the normalization of the H1 data within 
large theoretical uncertainty, although fails to describe the shape of the elasticity distribution. 
At HERA the $J/\psi$ polarisation parameters, extracted using the angular measurements, might 
indicate a need for CO terms in the QCD calculations. Various predictions show different deficits to 
describe the precise HERA data, and apparently higher order calculations would be necessary to draw 
a final conclusion.
        
\begin{figure}[htbp]
\centerline{\resizebox{0.54\textwidth}{!}{\includegraphics{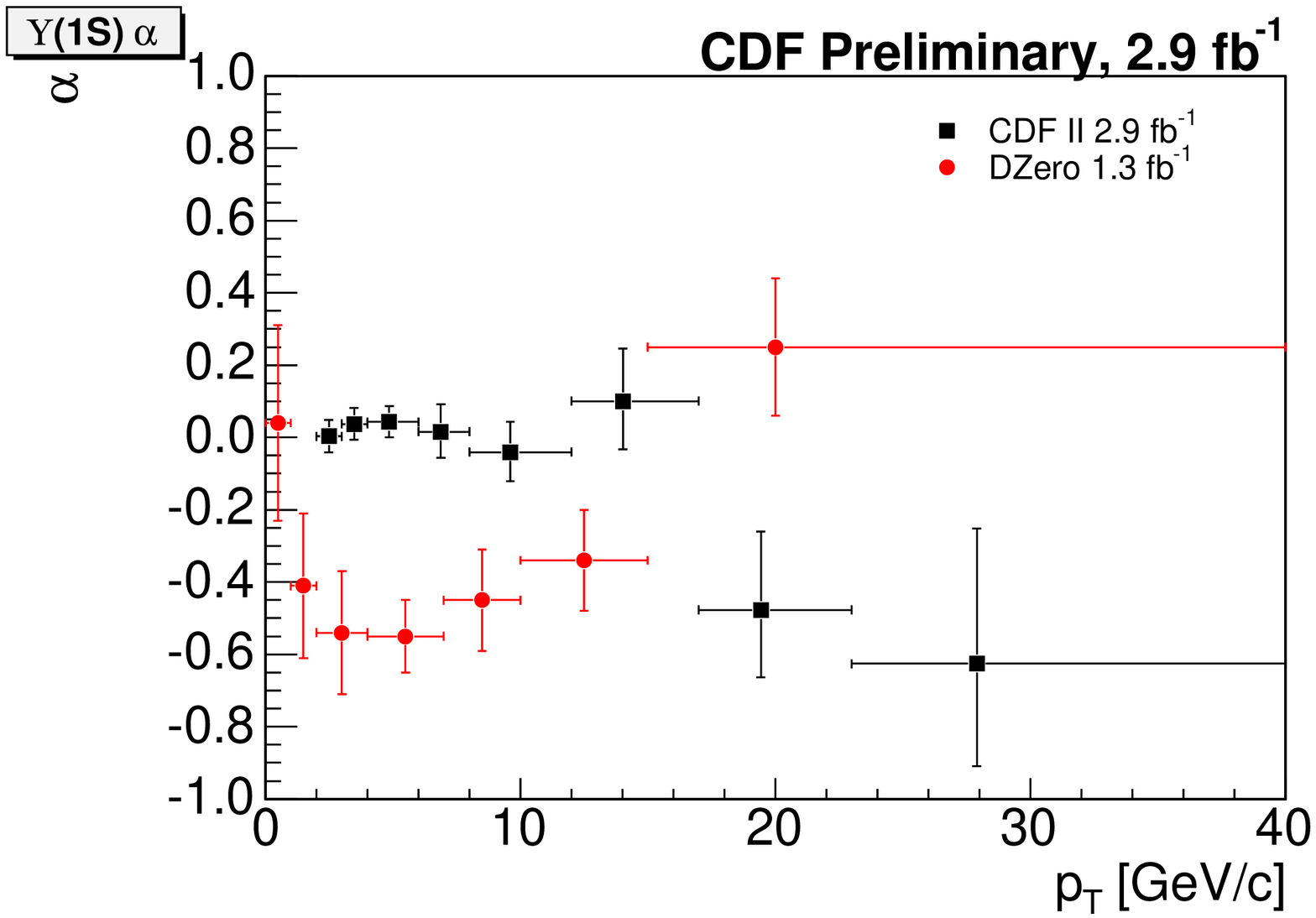}}%
\hfill%
\resizebox{0.54\textwidth}{!}{\includegraphics{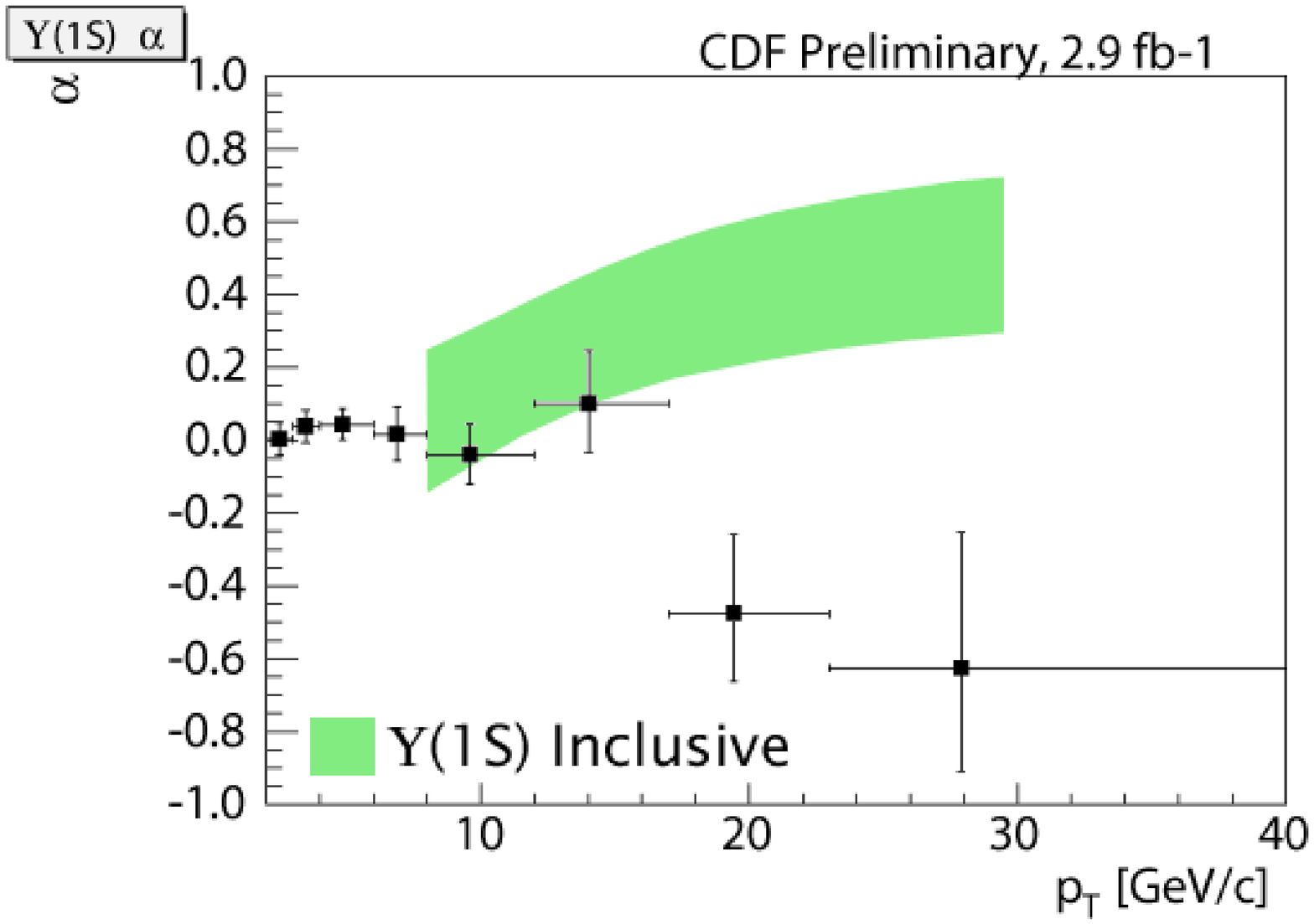}}}
\caption{Left: Polarisation parameter  for  the
$\Upsilon(1S)$ as a function of transverse momentum as measured by CDF and D0. 
Right: CDF data compared to the NRQCD prediction (shaded band). }
\label{fig2}
\end{figure}

Disagreement of $J/\psi$ polarisation measurement in $p\bar p$ collisions with NRQCD prediction was 
observed earlier at Tevatron~\cite{abe}. The CDF experiment presented a measurement of the 
$\Upsilon(1S)$ polarisation~\cite{kuhr} as shown in Fig.~\ref{fig2}. The preliminary data indicate a 
trend towards longitudinal polarisation and disagree with the NRQCD prediction. Also, a disagreement between 
CDF and D0 measurements was pointed out.

\subsection{Open charm and beauty production in QCD analysis of the nucleon structure}

At HERA open charm and beauty quarks are produced dominantly in boson-gluon fusion, a 
process particularly well suited to study QCD. 
Production of charm and beauty (jets) in DIS and photoproduction provides a powerful 
testing ground for pQCD calculations. Experimentally-determined charm and 
beauty structure functions set constraints on 
different phenomenological prescriptions for the treatment of heavy 
flavours in the QCD analyses of the proton structure. 
Furthermore, such measurements provide a direct cross check of the gluon distribution 
function obtained from scaling violations. At HERA charm and beauty events are tagged 
either in full reconstruction of heavy-flavoured mesons, or using the vertex-detector 
information, sensitive to the long lifetime of such hadrons. Results obtained by different 
methods can be combined. 

\subsubsection{QCD tests using open charm and beauty production at HERA}

\begin{figure}[htbp]
 \centerline{\resizebox{0.65\textwidth}{!}{\includegraphics{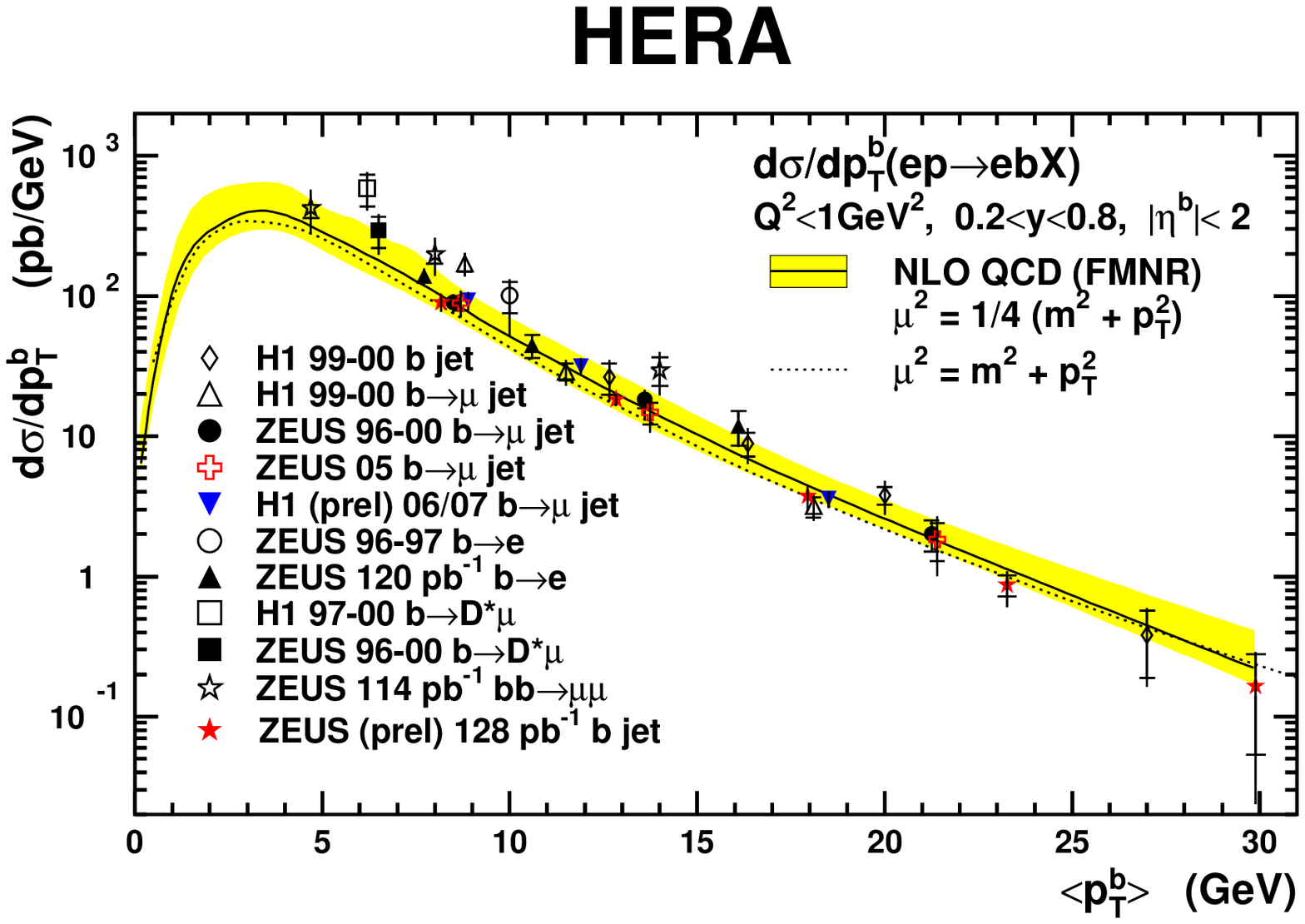}}%
\hfill%
\resizebox{0.44\textwidth}{!}{\includegraphics{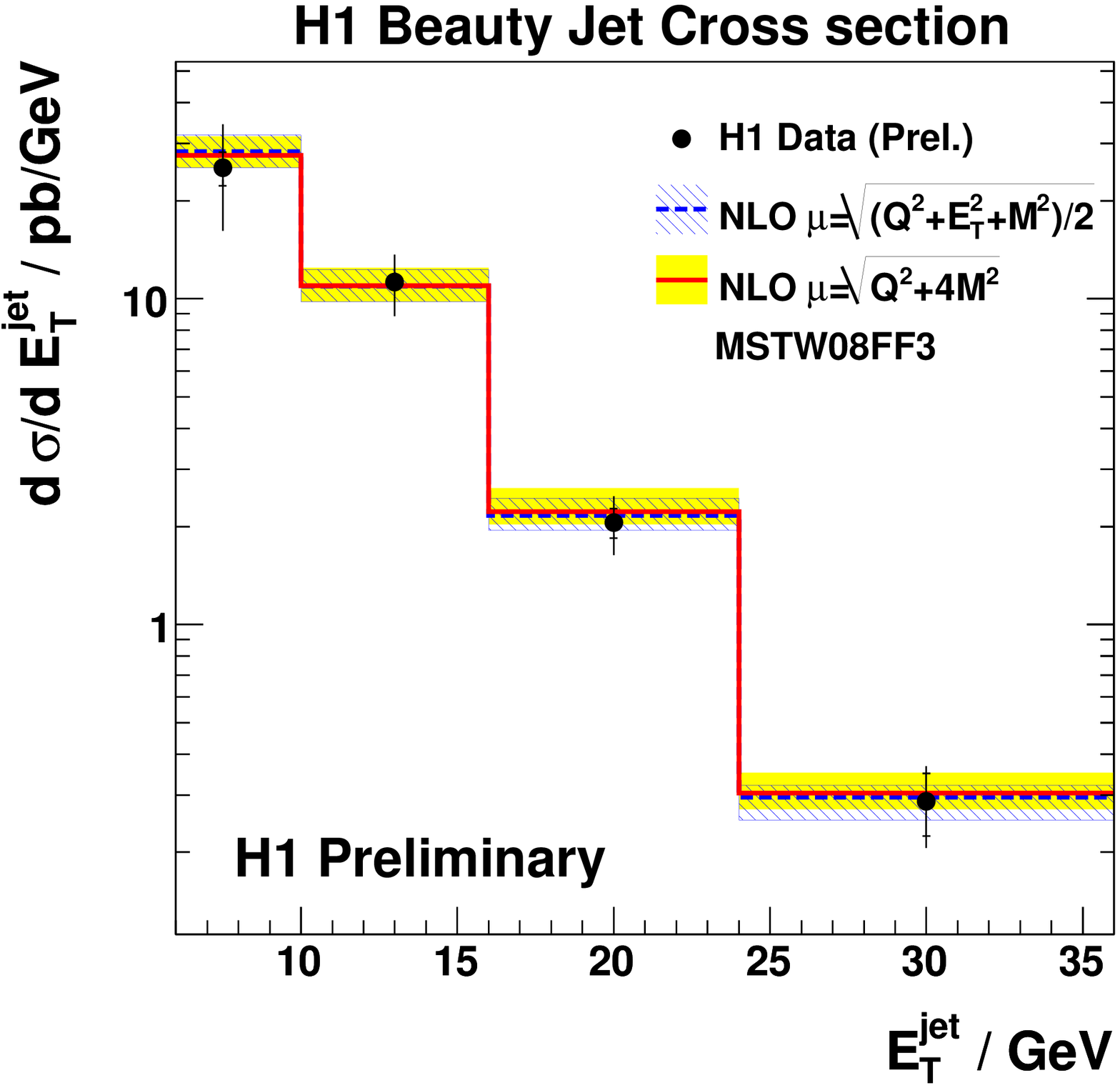}}}
\caption{ Left: HERA results on beauty photoproduction cross-section as a 
function of transverse momentum of the b-quark. Different symbols correspond to 
various beauty tagging methods. Measurements are compared to the NLO QCD (shaded band). 
Right: Beauty jet production cross-section in DIS as a function of transverse jet energy 
compared to the NLO QCD calculation for different scale choice (shaded bands).}
\label{fig3}
 
\end{figure}

Recent HERA results on beauty production in DIS~\cite{roloff,thompson} and in photoproduction~\cite{aushev} were discussed. 
In Fig.~\ref{fig3} HERA results on beauty production are shown in comparison with pQCD predictions at NLO 
in the fixed-flavour number scheme (FFNS), taking into account the $b$-quark mass. 
Different methods to tag beauty in photoproduction at HERA were used: the results 
agree very well and are described well by the NLO calculation. 
The cross-section of beauty jet production in DIS is measured as a function of the 
transverse jet energy and is described well by 
the massive NLO calculation even when using two different choices 
of the renormalisation and factorisation scales $\mu_r=\mu_f=\mu$. 
In contrary, charm-jet production in DIS is sensitive to the choice of 
the scales in the QCD calculation~\cite{thompson}.
 
\begin{figure}[htbp]
\centerline{\resizebox{0.52\textwidth}{!}{\includegraphics{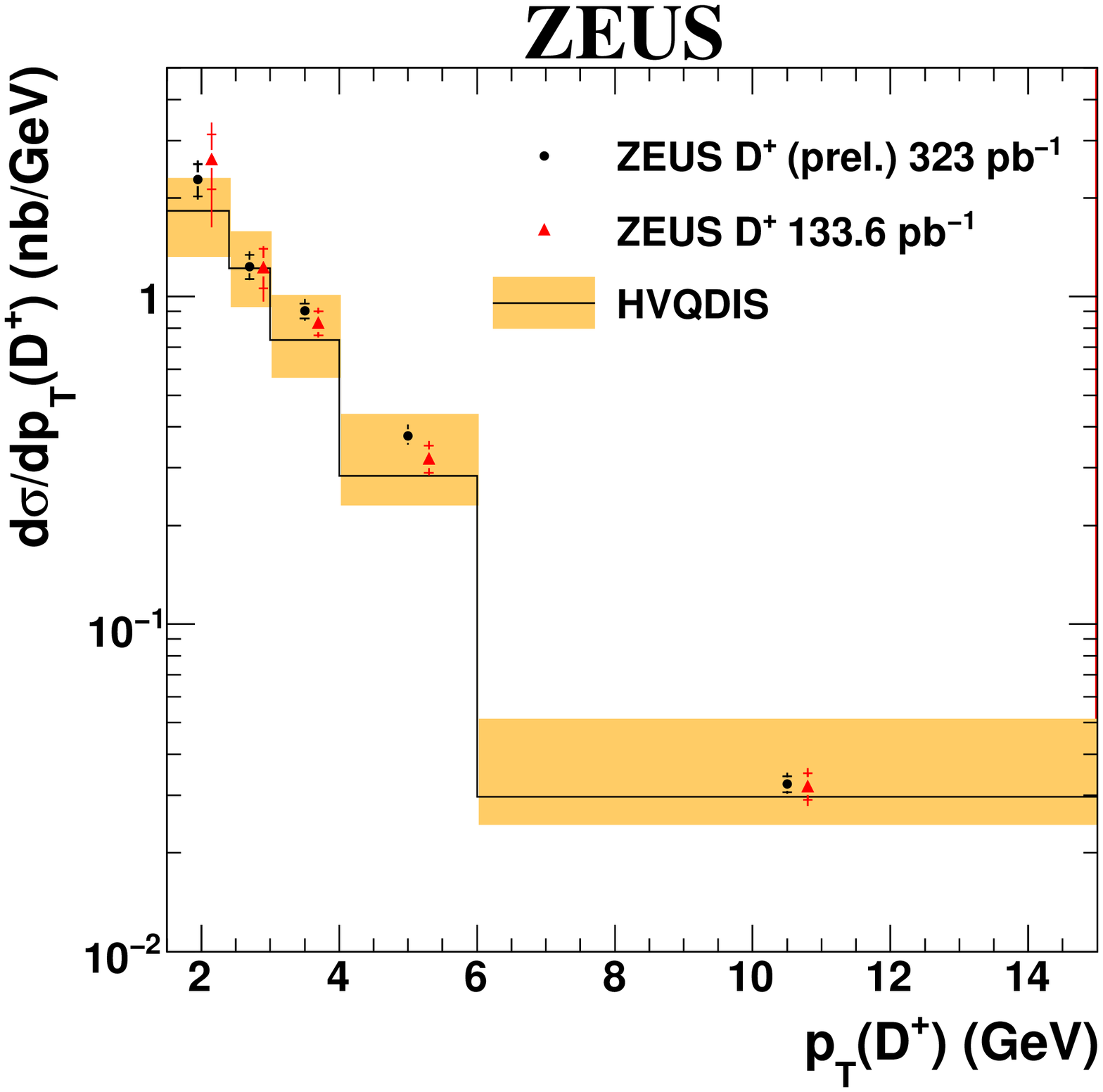}}%
\hfill%
\resizebox{0.52\textwidth}{!}{\includegraphics{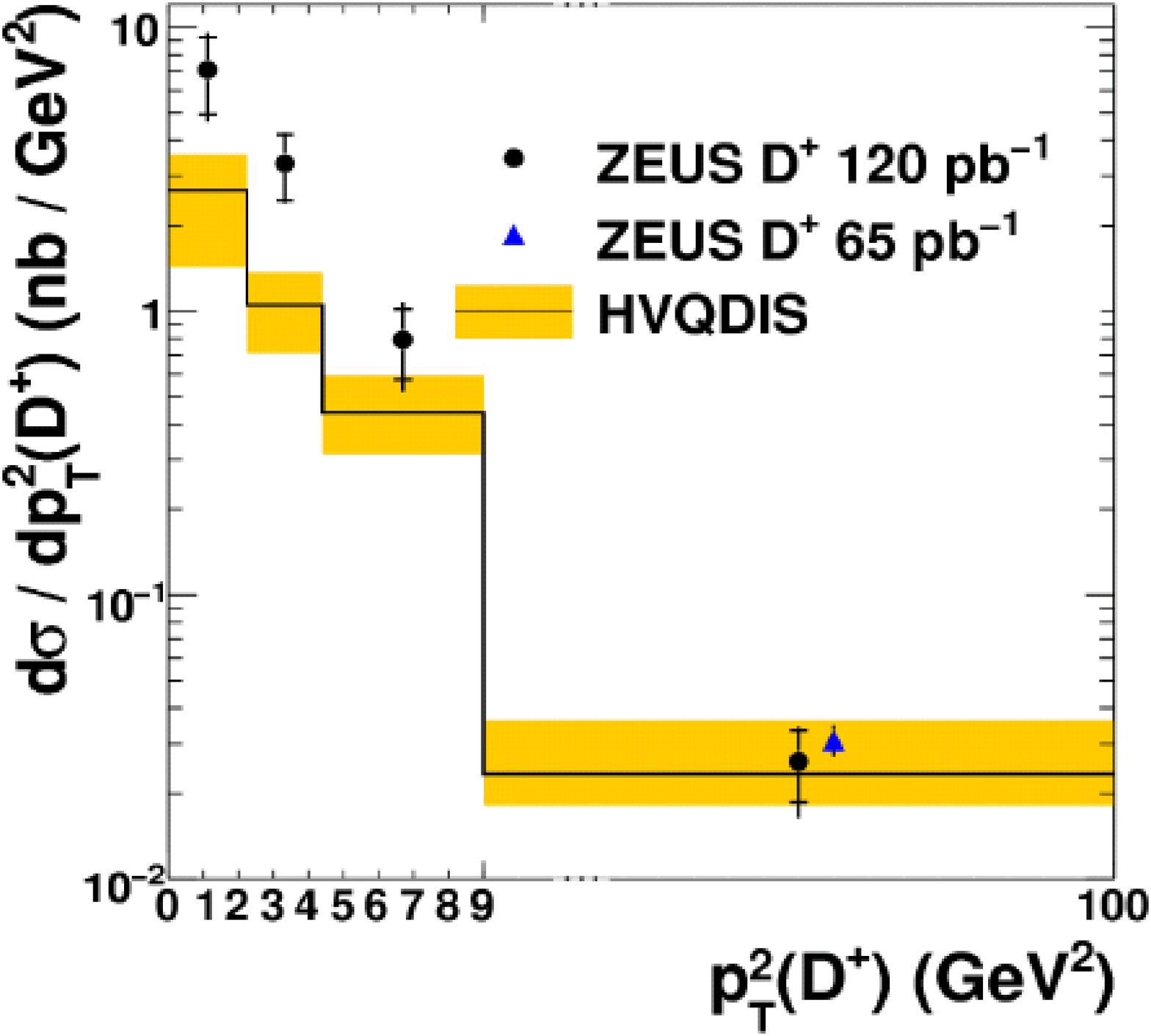}}} 
\caption{Left: Cross section of $D^+$ 
production in DIS as a function of the meson 
transverse momentum. Data (filled symbols) are compared to the 
NLO QCD (shaded band). Right: Measurement of $D^+$ 
production cross section at low transverse momenta as compared to the 
NLO QCD (shaded band).}
\label{fig4} 
\end{figure}

New analyses of open charm production in DIS~\cite{thompson,brinkmann,lisovy}
and in photoproduction~\cite{staykova} were presented. The cross sections of $D^+$-meson 
production are shown as a function of the meson transverse momentum in Fig.~\ref{fig4} 
and compared with the QCD calculation at NLO. Above the charm-production threshold (large $p_T(D^+)$) 
the calculation describes data well, while at threshold and below ($p_T(D^+)<1.5$~GeV) it underestimates the data.

\subsubsection{Heavy quark structure functions at HERA }
        
New measurements of the charm and beauty contributions, $F_2^c$ and $F_2^b$, 
to the inclusive proton structure function $F_2$ were presented~\cite{roloff,thompson,brinkmann,lisovy} and discussed in the 
context of QCD analyses of the proton structure. 

\begin{figure}[htbp]
\centerline{\resizebox{0.52\textwidth}{!}{\includegraphics{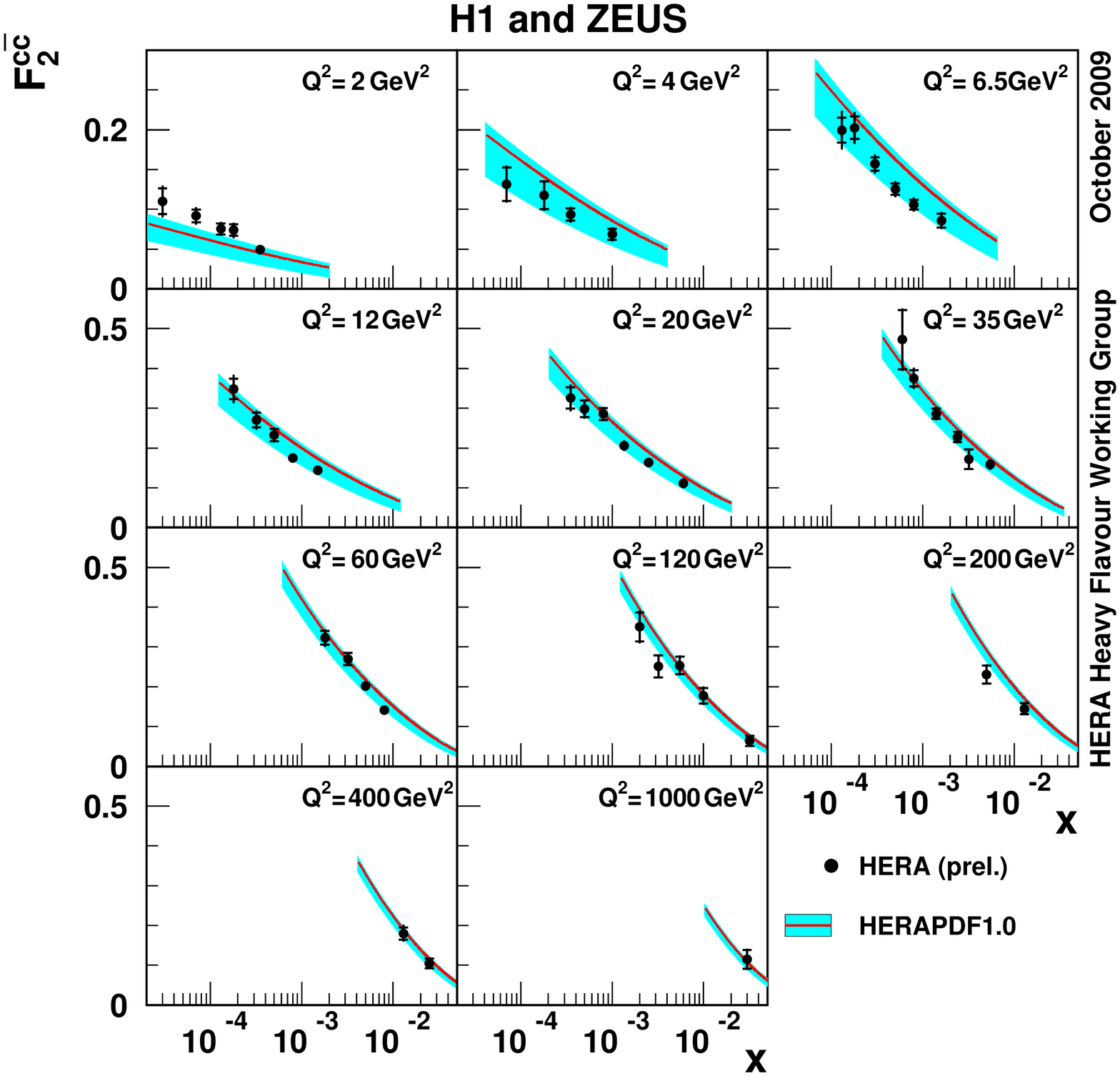}}%
\hfill%
\resizebox{0.55\textwidth}{!}{\includegraphics{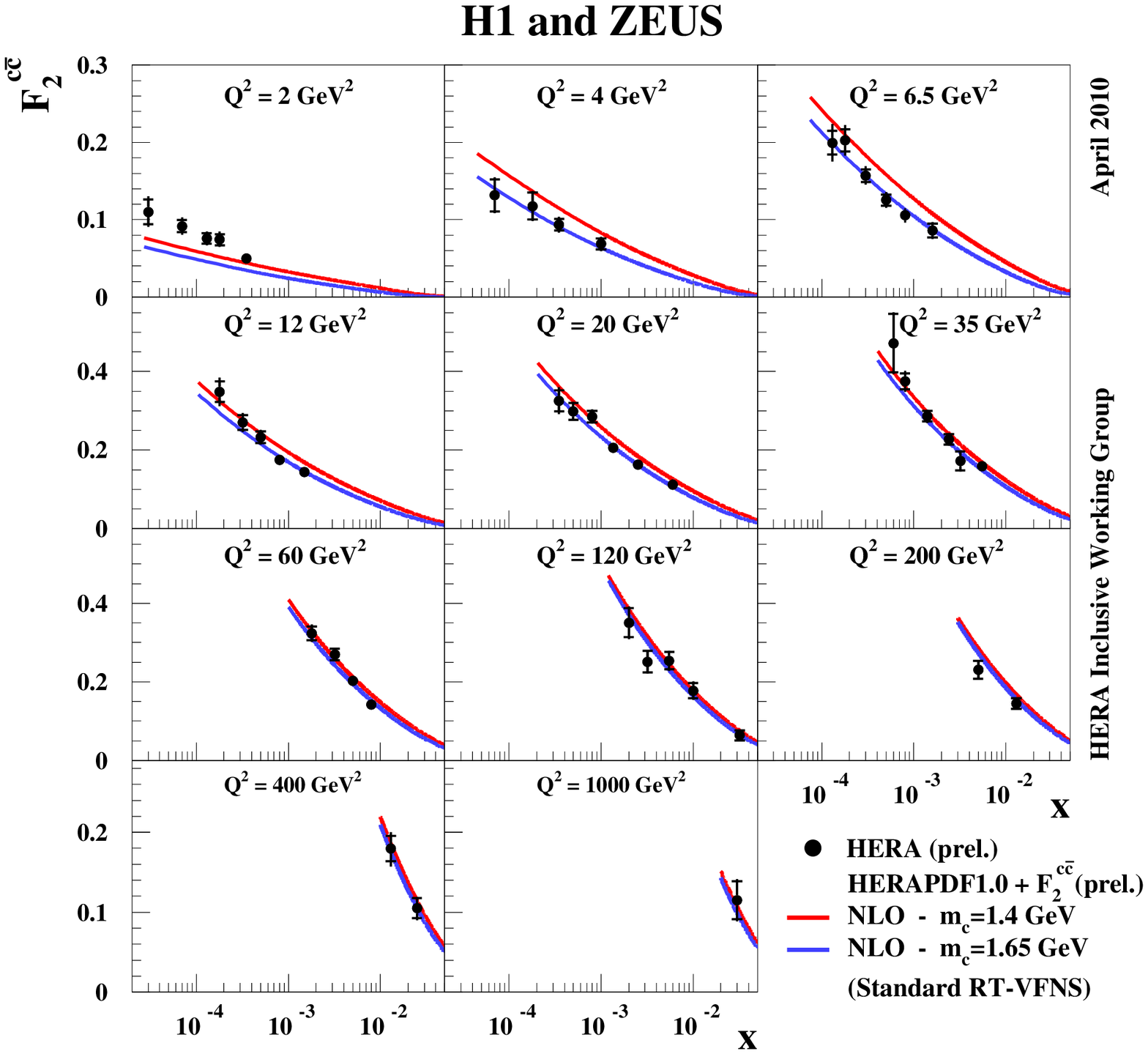}}}
\caption{Left: HERA combined $F_2^c$ 
as a function of Bjorken scaling variable $x$ in bins of photon 
virtuality Q2. The data (filled symbols) are compared to 
QCD prediction at NLO using HERAPDF1.0. Right:. The shaded band represent 
the uncertainty due to variation of charm mass in the PDF fit $1.35<m_c<1.65$~GeV. 
HERA $F_2^c$ as compared to QCD prediction using HERA PDF fit using 
$F_2^c$ data for $m_c=$~1.4~GeV (red line) and $m_c$=1.65~GeV (blue line).}
\label{fig5}
 \end{figure}

A topic of special interest was the HERA combined $F_2^c$ measurement~\cite{daum}, where various 
methods of charm tagging at both H1 and ZEUS experiments were combined, taking into account systematic correlations. 
Due to partially orthogonal uncertainty sources in the different tagging methods, 
a significant improvement in the overall precision of $F_2^c$ is observed. 
The data were compared to a variety of QCD predictions based on different prescriptions of 
the heavy-quark treatment in the PDF fits. 
The average precision of the data is about 10\% and is significant to distinguish 
between different models. In Fig.~\ref{fig5} the HERA $F_2^c$ structure
function is compared with the QCD prediction using the
HERAPDF1.0 set \cite{hera}. 
The prediction of $F_2^c$ 
based on HERAPDF1.0 
(which does not include charm data) describes the 
data very well except for low photon virtualities of $Q^2=2$~GeV$^2$. 
The variation of the charm mass $m_c$, 1.35~GeV$<m_c<1.65$~GeV,
in the PDF fit is investigated and accounted for as an additional model 
uncertainty in HERAPDF1.0. 
The choice of a larger $m_c$ was motivated by a recent determination of the charm 
pole mass. The $F_2^c$ measurements appear to be very sensitive to the choice of $m_c$ and 
exhibit a clear preference for the larger values of $m_c$. Choice of large $m_c$ in the QCD 
analysis of inclusive data results in a steeper gluon distribution. A sizable effect on the light quark 
distributions is also observed. These effects are not only very important for the QCD tests at HERA. Such, 
the NLO predictions for the cross-sections of $W$- and $Z$-boson production at the LHC using the 
PDFs determined with $m_c=1.65$~GeV rise by 3\% compared to those using PDFs with $m_c=1.4$~GeV.
A new PDF fit~\cite{mandy} using the preliminary charm data was performed at HERA for $Q^2>3.5$~GeV$^2$, 
using the same formalism as for HERAPDF1.0. Two values of $m_c$ were studied, resulting in a best fit 
for $m_c=1.65$~GeV and a steeper gluon distribution. Different prescriptions of the heavy-quark 
treatment in the PDF analyses were tested using HERA charm data. High sensitivity of the charm 
measurements to the heavy-flavour schemes in the PDFs in demonstrated.
However, the effects due to the heavy-flavour scheme and the choice of the heavy-quark mass in the 
QCD analyses are difficult to disentangle, and it is indeed impossible with inclusive DIS data alone. 
Therefore, heavy-flavour measurements at HERA provide a unique opportunity to test the heavy-flavour 
treatment schemes, however using an externally determined value of charm quark mass.

\subsubsection{Charm structure function in neutrino-nucleon scattering}

\begin{figure}[htbp]
\centerline{\resizebox{0.59\textwidth}{!}{\includegraphics{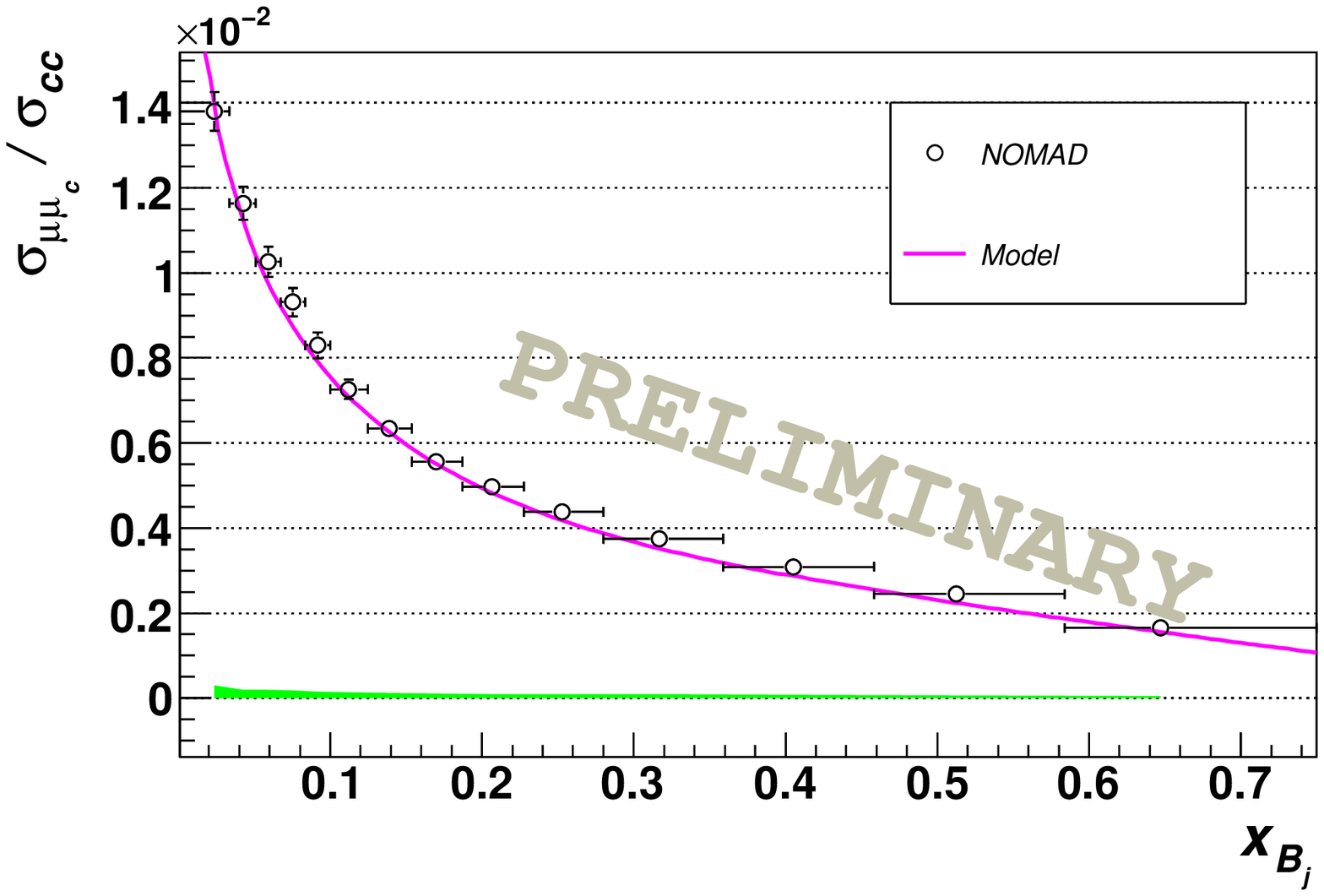}}%
\hfill%
\resizebox{0.44\textwidth}{!}{\includegraphics{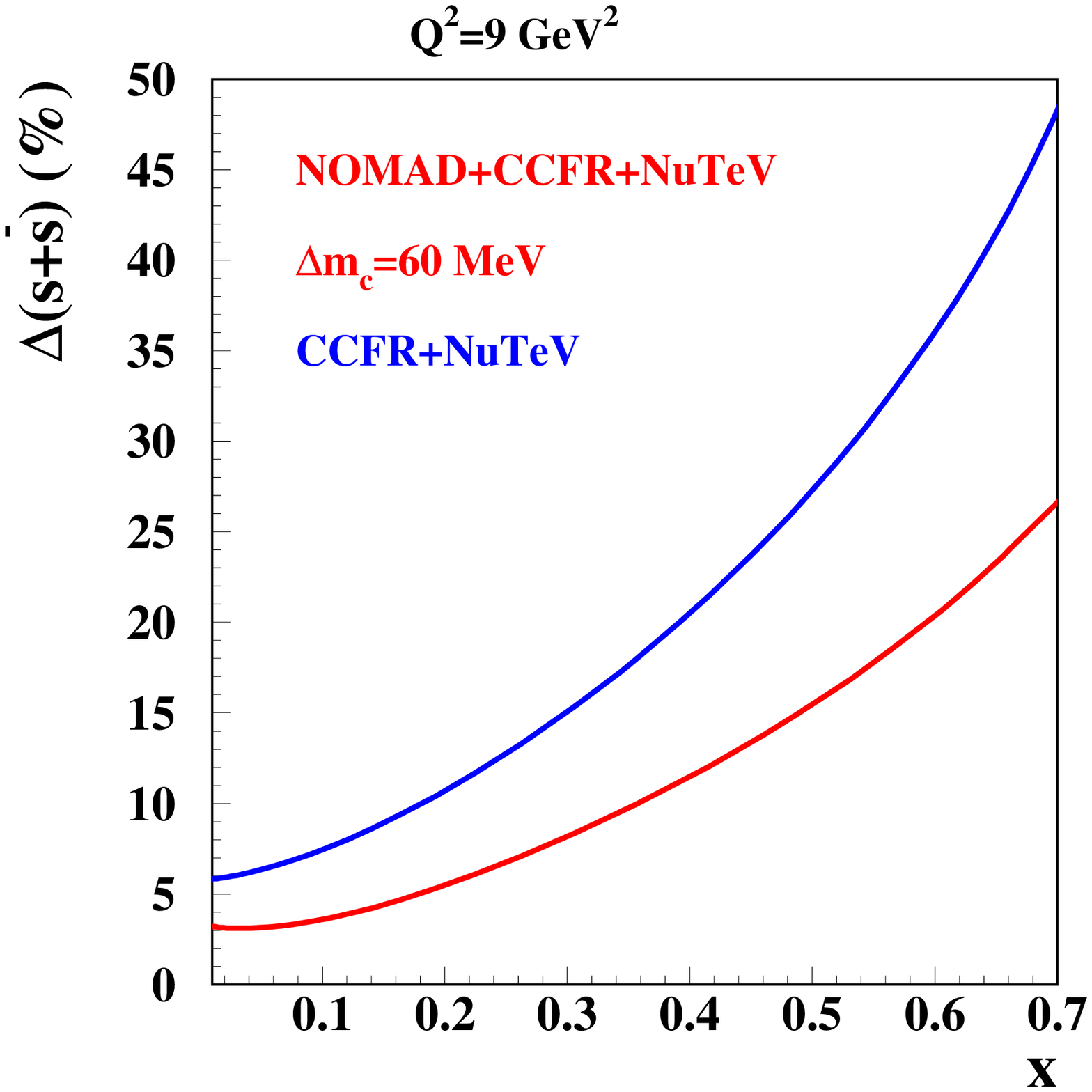}}} 
\caption{Left: Cross section ratio of charm production in dimuon-channel to the 
total charm cross section as a function of the
Bjorken scaling variable $x$. The data (open symbols) are compared to the QCD prediction (line). Right: Uncertainty 
on the strangeness distribution resulting from the PDF Fit before (blue) and after (red) including the NOMAD data.}
\label{fig6}
 \end{figure}

In neutrino-nucleon scattering the charm contribution $F_2^c$ 
to the inclusive structure function $F_2$ 
is sensitive to the strangeness distribution in the nucleon. 
The NOMAD collaboration presented impressive 
measurements of the ratio of di-muon to charm cross-sections close 
to the charm threshold~\cite{nomad}. 
In Fig.~\ref{fig6} the cross-section ratio is shown as a function of Bjorken variable 
$x$. The data have a significant impact on the QCD analyses of the proton structure. Using NOMAD 
measurements in PDF fits results in improvement of the precision on the strange sea distribution 
by a factor of 2.  

\subsection{Standard Model tests using heavy-quark production in $e^+e^-$ annihilation}

Measurements of charm and beauty production in $e^+e^-$ collisions play a crucial role in the understanding 
of the heavy-flavour sector of the Standard Model of electroweak interactions. Recent results on charm 
physics from Belle and BaBar were presented~\cite{milanes,kwon}, including studies of charm mixing, CP violation in the 
charm sector and properties of charmed meson decay. Measurements of the $D_s$ pseudoscalar purely leptonic 
decay branching fractions are also reported and a comparison with the lattice calculation of the 
$f_{D_s}$ decay constant is shown. The discrepancy of the current $f_{D_s}$ world average with the theoretical value 
is reduced. Recent results on charm mixing and CP-violation in $D^0$ and 
$\bar D^0$ are presented: the mixing parameters are in a good agreement with the Standard Model expectations, 
whereas no evidence for T-violation in $D^0$ multi-body decays is found. BaBar and Belle report disagreement between 
their measurements of polarization in $B$-meson decays to vector final states. Moreover, for the first time an 
observation of 4-body charmless baryonic $B$ decays was presented. 

\subsection{Charm and beauty in heavy ion collisions}
 
At RHIC heavy quarks are used as a probe of quark-gluon plasma. STAR and PHENIX
reported on recent results on charm and beauty production \cite{graz,xie}.
Heavy-quark production is tagged in semi-leptonic decays into electrons. 
After the new analysis of the STAR data, the electron cross-section measurements 
by the RHIC experiments exhibit very good agreement and are well described by the 
FONLL prediction. 
Future plans for heavy-quark physics for RHIC-II and first steps towards 
physics analysis at the ALICE were also presented \cite{rossi}.
  
\subsection{Top-quark production and properties}

\begin{figure}[htbp]
\centerline{\resizebox{0.49\textwidth}{!}{\includegraphics{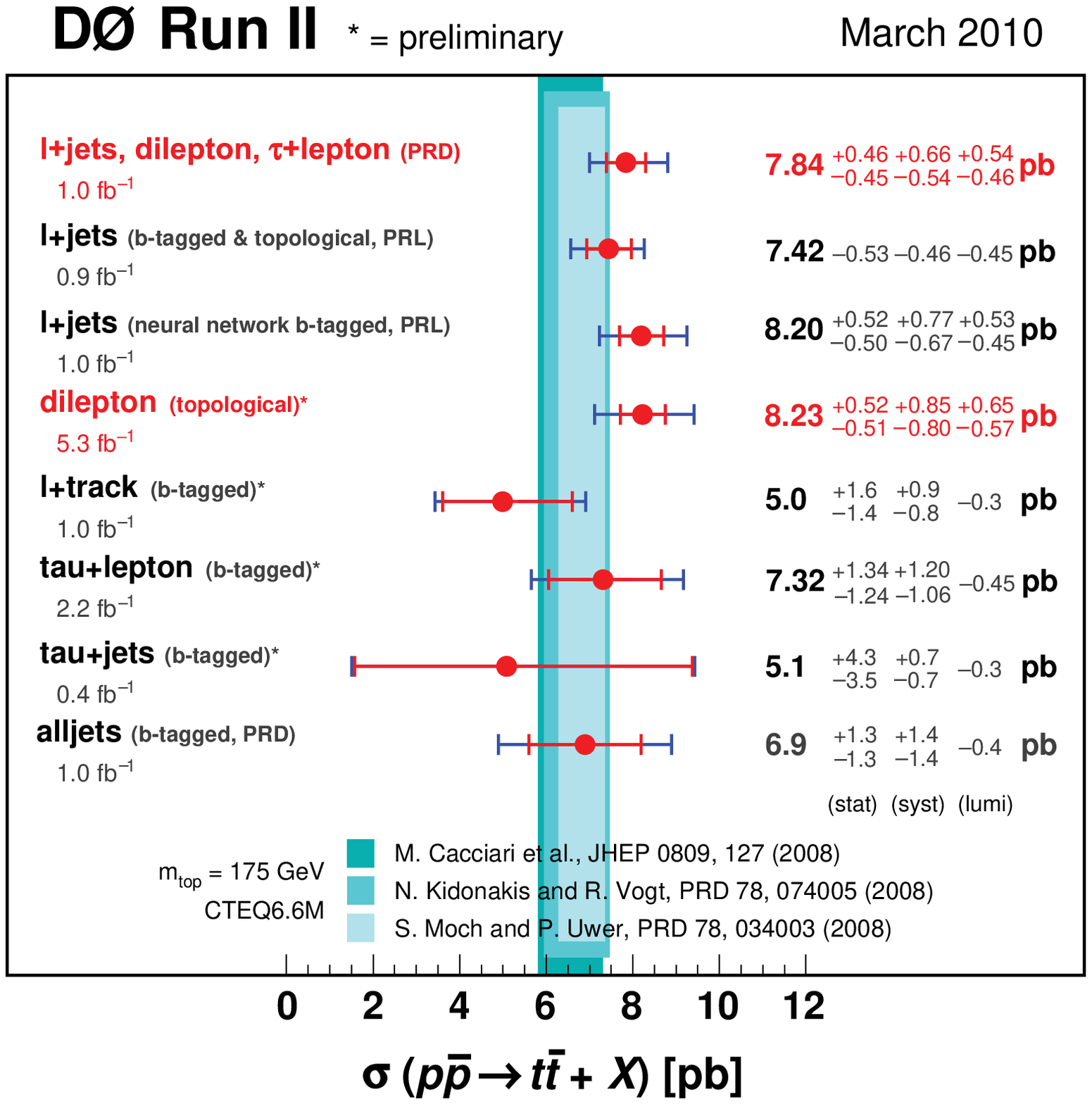}}%
\hfill%
\resizebox{0.42\textwidth}{!}{\includegraphics{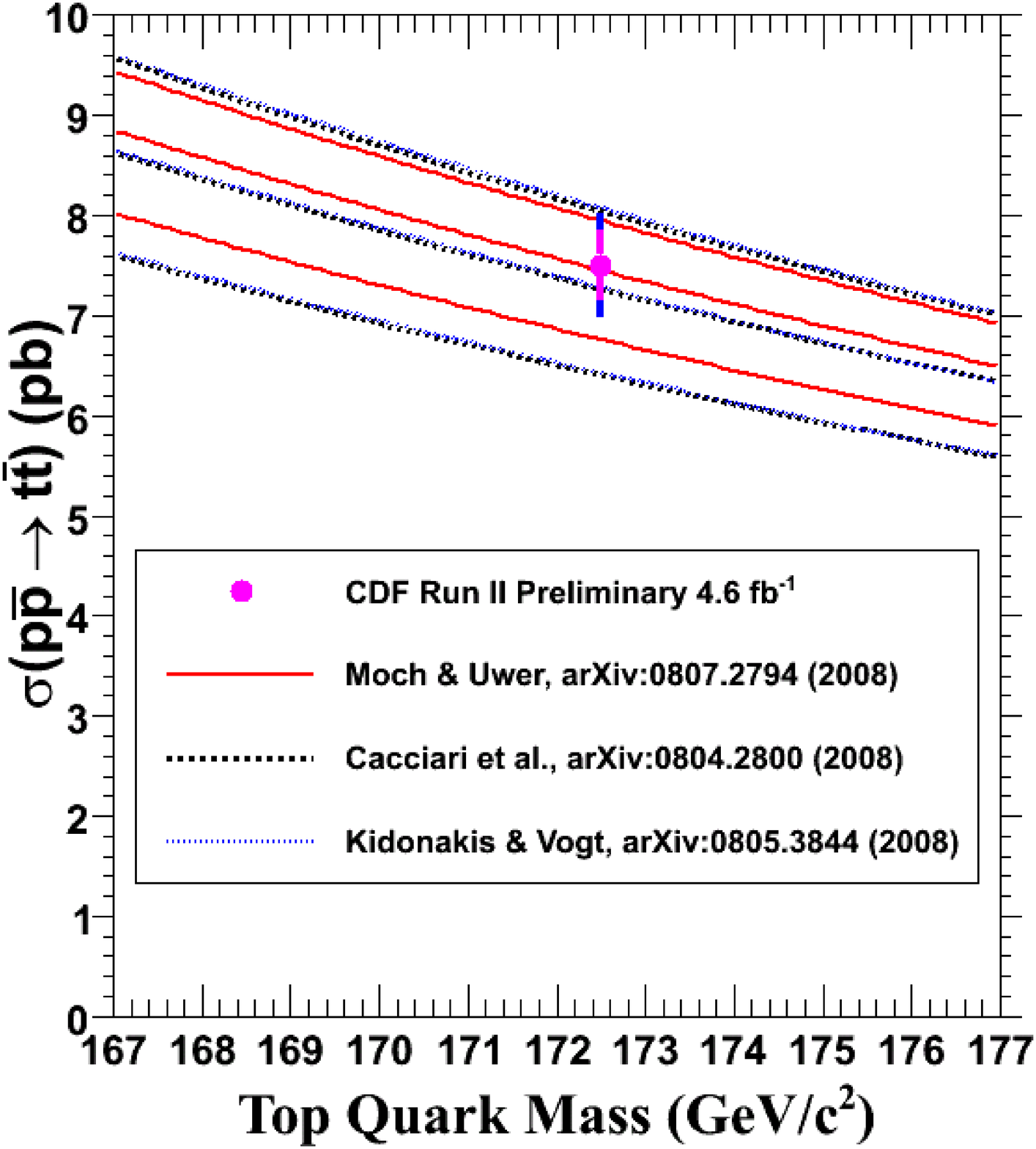}}} 
\caption{Left: Cross section of top-pair production as measured by D0 experiment 
in different channels of W decay. The data (filled symbols) 
are compared with a few calculations,
as discussed in the text. 
Right: The cross-section of top-pair production as 
a function of a top quark mass compared to the theoretical predictions. }
\label{fig7}
\end{figure}

Due to its very large mass, the top quark does not hadronize, but decays into
a $W$ and a $b$ quark, with a branching ratio of almost 100\%.
Such, an insight to the `bare' quark properties is possible. 
Precision measurements of the 
top-quark properties provide precision tests of the Standard 
Model. Fig.~\ref{fig7} shows 
the cross section measurements of top-pair production at the 
Tevatron \cite{thery,compostella}, 
compared with theoretical predictions at NLO, with the possible
inclusion of NLL soft/collinear resummation 
and some NNLO contributions. 
Several analysis methods 
are explored and all the top decay channels 
are considered, in order to better constrain the properties of 
the top quark and to search for possible sources 
of new physics affecting the $t\bar t$-production mechanism. 

Precision measurements of the cross 
section can also be used for the 
extraction of the pole top quark mass using higher-order calculations.
However, the theoretical uncertainties 
on the top cross section predictions 
are quite large. Direct measurement of the top quark mass 
using various methods by CDF and D0 collaborations also discussed~\cite{adelman,petrillo}.
These measurements can reach very high precision, but are 
limited by systematic uncertainty. 
However, since such analyses are mostly driven by Monte Carlo event generators,
the physical interpretation of the measured top mass in terms of 
a well-defined theoretical definition is not straightforward. 
Other top quark properties, such as charge, helicity, 
width and lifetime have been measured at the Tevatron~\cite{park} and have been shown 
to be consistent with the Standard Model expectations.
Ref.~\cite{donini} presents the
plans and strategies of ATLAS and CMS to measure the $t\bar t$ cross section
using early LHC data and a centre-of-mass energy of 10~TeV.
According to \cite{donini}, it will be possible to measure this
cross section with a precision of 20-30\% even with a luminosity 
of a few tens inverse picobarns.

\section{Conclusions}
The session on heavy flavours in DIS and hadron colliders at the
DIS2010 workshop had a number of impressing presentations,
discussing recent theoretical improvements as well as experimental results
and prospectives for heavy-quark and heavy-hadron production in 
several different experimental environments.
Especially worth 
to mention are the latest calculations in perturbative QCD, NRQCD, Regge 
and non-perturbative models, as well as presentations of heavy-ion collisions, 
developments of Monte Carlo generators and interpretation of exotic hadrons observed at the Tevatron.
Latest experimental results from major experiments as HERA, Tevatron, RHIC and $B$-factories, as well
as the prospects for heavy-flavour measurements at the LHC were presented.
In general, the experimental data are in reasonable agreement with 
the theoretical predictions, however, as discussed throughout this paper,
there are still several open issues calling for more refined
computations, along with more precise data. With the increasing precision of the measurements, 
close collaboration between experimentalists and theorists will be 
inevitable for probing 
and understanding the mechanisms of heavy-flavour production and 
its particular role in the structure of matter.

\end{document}